\documentclass[prd,nofootinbib,amsfonts,notitlepage]{revtex4-1}
\usepackage[utf8]{inputenc}
\usepackage[T1]{fontenc}
\usepackage{hyperref}
\usepackage{graphicx,bm,amsmath,color}
\usepackage{bbold}
\usepackage{wasysym}
\usepackage{verbatim}
\usepackage{amssymb}
\usepackage{epstopdf}
\usepackage{animate}
\usepackage{xmpmulti}
\usepackage{centernot}
\usepackage{multirow}
\usepackage{tikz}
\usepackage{graphicx}
\usepackage{float}
\usepackage{mathtools}
\usepackage{comment}
\usetikzlibrary{arrows}
\usetikzlibrary{decorations.pathreplacing}

\makeatletter

\newcommand{\be}{\begin{equation}}
\newcommand{\ee}{\end{equation}}
\newcommand{\beq}{\begin{eqnarray}}
\newcommand{\eeq}{\end{eqnarray}}
\newcommand{\ba}{\begin{align}}
\newcommand{\ea}{\end{align}}

\begin{document}

\title{Black hole surface gravity in doubly special relativity geometries}
\author{José Javier Relancio}
\email{relancio@unizar.es}
\affiliation{Dipartimento di Fisica ``Ettore Pancini'', Università di Napoli Federico II, Napoli 80138, Italy;\\
INFN, Sezione di Napoli, Napoli 80126, Italy;\\
Centro de Astropartículas y Física de Altas Energías (CAPA), Universidad de Zaragoza, Zaragoza 50009, Spain}

\author{Stefano Liberati}
\email{liberati@sissa.it}
\affiliation{SISSA, Via Bonomea 265, 34136 Trieste, Italy and INFN, Sezione di Trieste;\\ IFPU - Institute for Fundamental Physics of the Universe, Via Beirut 2, 34014 Trieste, Italy}

\begin{abstract}
In a quantum gravity theory, spacetime at mesoscopic scales can acquire a novel structure very different from the classical concept of general relativity. A way to effectively characterize the quantum nature of spacetime is through a momentum dependent space-time metric. There is a vast literature showing that this geometry is related to deformed relativistic kinematics, which is precisely a way to capture residual effects of a quantum gravity theory. In this work, we study the notion of surface gravity in a  momentum dependent Schwarzschild black hole geometry. We show that using the two main notions of  surface gravity in general relativity we obtain a momentum independent result. However, there are several definitions of surface gravity, all of them equivalent in general relativity when there is a Killing horizon. We show that in our scheme,  despite the persistence of a Killing horizon, these alternative notions only agree in a very particular momentum basis, obtained in a previous work, so further supporting its physical relevance.
\end{abstract}

\maketitle

\section{Introduction}
 It is common lore that the most daunting challenge of theoretical physics is nowadays the unification of General Relativity (GR) and Quantum Field Theory (QFT), or equivalently, the formulation of a Quantum Gravity Theory (QGT). Indeed, while GR and QFT had a stunning success in describing the observed natural phenomena, they also showed fundamental incompatibilities, mostly stemming from the role that spacetime plays in them (a dynamical variable in GR, a static frame in QFT).

Such apparent incompatibility is paradigmatically illustrated by the so-called information loss problem associated to quantum black hole evaporation~\cite{Hawking:1976ra}. On the one hand, accepting the standard scenario dictated by GR seems in fact to imply an non-unitary evolution of quantum states, so violating a basic tenet of QFT. On the other hand avoiding this loss of unitarity  seems to  require a radical departure from the equivalence principle that would predict no radical departure from standard physics at the event horizon of the black hole (see e.g. the so-called firewall paradigm proposed in~\cite{Almheiri:2012rt}).

In order to solve these inconsistencies between general relativity and quantum field theory, several QGT proposals have been advanced and developed in the past decades. Examples of these attempts are string theory~\cite{Mukhi:2011zz,Aharony:1999ks,Dienes:1996du}, loop quantum gravity~\cite{Sahlmann:2010zf,Dupuis:2012yw}, 
causal dynamical triangulations \cite{Loll_2019} or causal set \mbox{theory~\cite{Wallden:2013kka,Wallden:2010sh,Henson:2006kf}}. In most of these theories, a minimum length appears~\cite{Gross:1987ar,Amati:1988tn,Garay1995}, which is normally associated with the Planck length $\ell_P \sim 1.6\times 10^{-33}$\,cm. It is believed that this minimum length could mark somehow the transition to a ``quantum'' spacetime which replaces our concept of ``classical'' spacetime. Unfortunately, the aforementioned theories are not yet fully satisfactory in the sense that they do not have yet well defined testable predictions which might serve us as a guidance in building a definitive theory of quantum gravity.

However, a complementary approach toward the realization of a quantum gravity theory can be represented by a bottom up strategy where the possible scenarios of a low-energy (sub-Planck scale) limit of quantum gravity are considered and put to the observational/experimental tests.
In particular, it is expected that the transition from a full quantum and discrete spacetime to the standard classical continuum one, will not happen abruptly but will give rise to a mesoscopic regime where a continuum spacetime is endowed with different local symmetries due to remnant structure inherited from the super-Planckian regime. Deviation from standard Lorentz invariance is the most investigated scenario (but not the only one, see e.g.~\cite{Belenchia:2015ake}). In this respect there are two main scenarios: one can consider that for high energies a Lorentz invariance violation (LIV)~\cite{Colladay:1998fq,Kostelecky:2008ts} can arise, or that the symmetry is deformed, leading to the theories known as deformed special relativity (DSR)~\cite{AmelinoCamelia:2008qg}.

LIV scenarios modify the kinematics of special relativity (SR) with the introduction of a preferred frame associated to some extra geometrical structure such as a fixed norm vector field. Such framework allows then to write for elementary particles a modified, no more Lorentz invariant, dispersion relation. Generally,  new terms proportional to the inverse of a high-energy scale (normally considered to be the Planck scale) are added to the usual quadratic expression of SR. 

In DSR theories the relativity principle is instead preserved, albeit at the cost of introducing a non-linear realization of the Lorentz group which allows for an invariant (observer independent) energy scale. Also in this case, this quantum gravity scale can be associated to a deformed dispersion relation, although in this framework this is not fully capturing the new physics. Indeed, one can even choose a special basis in momentum space, the so-called ``classical basis'' of $\kappa$-Poincaré~\cite{Borowiec2010}, where the usual dispersion relation of SR is recovered. However, even in this case one gets that the deformed symmetry requires a deformed composition law for the momenta. This implies that the total momentum of a system of two (or more) particles is not derivable as the trivial sum of the initial momenta as in SR (it involves instead additional terms depending on both momenta and on the high-energy scale).

Also in this context black holes have been proved to be the best objects for testing new scenarios in gravitational physics. The study of black holes in LIV frameworks was considered in~\cite{Dubovsky:2006vk,Barausse:2011pu,Blas:2011ni,Bhattacharyya:2015gwa}. In these works it was shown that particles with different energies see different horizons. Moreover, the fact that generically there is not a common Killing horizon in LIV theories is problematic as it seems that black holes could violate the second law of thermodynamics already at the classical level as well as via particle dependent Hawking radiation (see e.g.~\cite{Dubovsky:2006vk}), 
albeit the dynamical realization of such violations might be precluded (see e.g.~\cite{Benkel:2018abt}). 
The UV completion of LIV theories could remedy at this problem by introducing further geometrical structure to the black holes (the so-called Universal Horizon~\cite{Blas:2011ni}) which might fix a universal temperature and restore black hole thermodynamics~\cite{Herrero-Valea:2020fqa}.

In DSR scenarios a natural question concerns the possibility for then to admit a geometrical description, and it was soon understood that a momentum dependence of spacetime would naturally arise. Indeed, in~\cite{Carmona:2019fwf} it was rigorously shown that all the ingredients of a relativistic deformed kinematics can be obtained from a maximally symmetric momentum space. In particular,   $\kappa$-Poincaré kinematics can be obtained identifying the isometries (translations and Lorentz isometries) and the squared distance of the metric with the deformed composition law, deformed Lorentz transformations and deformed dispersion relation respectively (the last two facts were previously contemplated in Refs.~\cite{AmelinoCamelia:2011bm,Lobo:2016blj}). In~\cite{Relancio:2020zok} the proposal of~\cite{Carmona:2019fwf} was generalized so allowing the metric to describe a curved spacetime, leading to a metric in the cotangent bundle depending on all the phase-space variables. This is a generalization of previous works in the literature, in which a metric that depends on the velocities (Finsler geometries)~\cite{Girelli:2006fw,Amelino-Camelia:2014rga,Lobo:2016xzq} and momenta (Hamilton geometries)~\cite{Barcaroli:2015xda,Barcaroli:2016yrl,Barcaroli:2017gvg} were regarded.

In the context of rainbow geometries (not directly linked to DSR), the study of the Hawking radiation of Schwarzschild black holes in the presence of momentum corrections of the black hole metric, and then on the dispersion relation, was studied in~\cite{Peng:2007nj,Ali:2014xqa}. In these papers it was considered that the energy of the modified dispersion relation was the mass of the black hole. In another vein, in~\cite{Li:2008gs} the momentum dependency arises from the energy of the particle emitted by the Hawking radiation. Moreover, the energy scale of the modification of  Hawking radiation was taken to be the inverse of the Schwarzschild radius in~\cite{Gim:2014ira,Gim:2015yxa,Mu:2015qna,Kim:2016qtp,Tao:2016baz,Bezerra:2017hrb,Feng:2017iso,Feng:2018jqf,Shahjalal:2018hid}. In all the previous scenarios, this could lead to a remnant mass, i.e., the black hole cannot fully evaporate. Noticeably, in~\cite{Yadav:2016nfh} it was studied the Unruh effect, obtaining a similar modification found in the previous papers.

In a previous work these authors have advanced a proposal for a geometrical description of DSR in curved spacetimes via cotangent bundle geometries \cite{Relancio:2020zok,Relancio:2020rys,Relancio:2020mpa}. Black holes were considered in~\cite{Relancio:2020zok} where a common horizon for all particles, independently of their energy, was shown to exists. 

In this paper we focus on the notion of surface gravity in the DSR scenario, and in particular, in the geometrical interpretation of the latter considered in Refs.~\cite{Relancio:2020zok,Relancio:2020mpa,Relancio:2020rys,Pfeifer:2021tas}. In GR there is a simple geometrical way to relate the surface gravity to Hawking radiation. However, the rigorous way in which one is able to obtain this phenomenon is by a QFT in curved spacetimes~\cite{Birrell:1982ix,Wald:1984rg}. Then, in this context we will propose a way to define Hawking temperature in the same line of the above mentioned works on rainbow geometries, but the only way in which this can be derived consistently is by a DSR QFT. While there are some works on this topic~\cite{Kosinski:2001ii,Govindarajan:2009wt,Poulain:2018two,Arzano:2020jro,Lizzi:2021rlb}, such a theory is far from being reached.

While, from both a geometrical and algebraical point of view, different choices of the kinematics of $\kappa$-Poincaré (different choices of coordinates in a de Sitter momentum space or different bases in Hopf algebras~\cite{KowalskiGlikman:2002we}) represent the same deformed kinematics (with the same properties, such as the associativity of the composition law and the relativity principle), there is an ambiguity about what are the momentum variables associated to physical measurements. The fact that different bases could represent different physics was deeply considered in the literature~\cite{AmelinoCamelia:2010pd}. In~\cite{Pfeifer:2021tas} we showed that only Lorentz covariant metrics are allowed in our geometrical scheme. Moreover, in~\cite{Relancio:2020rys} we proposed a way to select this ``physical'' basis by imposing the conservation of the Einstein tensor. Here, we will see that only for this particular basis all the definitions of surface gravity which coincide in GR also do in this context. 

The structure of the paper is as follows. We start by summarizing the concepts of the cotangent bundle geometry we use in the following in Sec.~\ref{sec:geometrical_intro}, where we also briefly discuss our main results of previous works. All the notions of surface gravity coincide in GR if Killing equation is satisfied. However, we find that only a particular momentum basis is able to do so in Sec.~\ref{sec:killing}.  In Sec.~\ref{sec:surface_gravity} we compute the two main notions of surface gravity of GR, the peeling off and inaffinity of null geodesics, showing that, in any basis allowed in our scheme, are always momentum independent, obtaining then the same result of GR. In Sec.~\ref{sec:killing_basis} we discuss that,  despite having a Killing horizon, different notions of surface gravity considered in GR lead to different (momentum dependent) results. The only way in which this can be avoided is by considering a particular momentum basis obtained in Sec.~\ref{sec:killing}. We check that for this preferred basis several notions agree in Sec.~\ref{sec:different_notions}. Finally, we end with the conclusions in Sec.~\ref{sec:conclusions}.   

\section{Cotangent bundle in a nutshell}
\label{sec:geometrical_intro}
In this section we review the main geometrical ingredients in the cotangent bundle approach that we shall use in the following. Also we shall recall the main results from our previous papers about how to consider a deformed relativistic kinematics in a curved space-time background.  

\subsection{Main properties of the geometry in the cotangent bundle}

In~\cite{miron2001geometry} a line element in the cotangent bundle is defined as  
\begin{equation}
\mathcal{G}\,=\, g_{\mu\nu}(x,k) dx^\mu dx^\nu+g^{\mu\nu}(x,k) \delta k_\mu \delta k_\nu\,,
\label{eq:line_element_ps} 
\end{equation}
where 
\begin{equation}
\delta k_\mu \,=\, d k_\mu - N_{\nu\mu}(x,k)\,dx^\nu\,. 
\end{equation}

In~\cite{miron2001geometry} it is shown that a horizontal curve {in the cotangent bundle} is determined by the geodesic motion in spacetime
\begin{equation}
\frac{d^2x^\mu}{d\tau^2}+{H^\mu}_{\nu\sigma}(x,k)\frac{dx^\nu}{d\tau}\frac{dx^\sigma}{d\tau}\,=\,0\,,
\label{eq:horizontal_geodesics_curve_definition}
\end{equation} 
and by the change of momentum obtained from
\begin{equation}
\frac{\delta k_\lambda}{d \tau}\,=\,\frac{dk_\lambda}{d\tau}-N_{\sigma\lambda} (x,k)\frac{dx^\sigma}{d\tau}\,=\,0\,,
\label{eq:horizontal_momenta}
\end{equation} 
where 
\begin{equation}
{H^\rho}_{\mu\nu}(x,k)\,=\,\frac{1}{2}g^{\rho\sigma}(x,k)\left(\frac{\delta g_{\sigma\nu}(x,k)}{\delta x^\mu} +\frac{\delta g_{\sigma\mu}(x,k)}{\delta  x^\nu} -\frac{\delta g_{\mu\nu}(x,k)}{\delta x^\sigma} \right)\,,
\label{eq:affine_connection_st}
\end{equation} 
is the affine connection of spacetime, and 
\begin{equation}
\frac{\delta}{\delta x^\mu}\, \doteq \,\frac{\partial}{\partial x^\mu}+N_{\nu\mu}(x,k)\frac{ \partial}{\partial k_\nu}\,.
\label{eq:delta_derivative}
\end{equation}
Here, $\tau$ plays the role of the proper time or the affine parameter depending if one is considering a massive or a massless particle respectively. 

The choice of the nonlinear connection coefficients $N_{\nu\mu}(x,k)$ is not unique but, as it is shown in~\cite{miron2001geometry}, there is one and only one choice of nonlinear connection coefficients that leads to a space-time affine connection which is metric compatible and torsion free.
In GR, the coefficients of the nonlinear connection are given by
\begin{equation}
N_{\mu\nu}(x,k)\, = \, k_\rho {\Gamma^\rho}_{\mu\nu}(x)\,,
\label{eq:nonlinear_connection}
\end{equation} 
where $ \Gamma^\rho_{\mu\nu}(x)$ is the affine connection. Then, when the metric does not depend on the space-time coordinates, these coefficients vanish.

In~\cite{miron2001geometry} it was defined the covariant derivatives in space-time 
\begin{equation}
\begin{split}
T^{\alpha_1 \ldots\alpha_r}_{\beta_1\ldots\beta_s;\mu}(x,k)\,&=\,\frac{\delta T^{\alpha_1 \ldots\alpha_r}_{\beta_1\ldots\beta_s}(x,k)}{\delta x^\mu}+T^{\lambda \alpha_2 \ldots\alpha_r}_{\beta_1\ldots\beta_s}(x,k){H^{\alpha_1}}_{\lambda \mu}(x,k)+\cdots+T^{\alpha_1 \ldots \lambda}_{\beta_1\ldots\beta_s}(x,k){H^{\alpha_r}}_{\lambda \mu}(x,k)\\
&-T^{\alpha_1 \ldots \alpha_r}_{\lambda \beta_2\ldots\beta_s}(x,k){H^{\lambda}}_{\beta_1 \mu}(x,k)-\cdots-T^{\alpha_1 \ldots \alpha_r}_{\beta_1\ldots \lambda}(x,k){H^{\lambda}}_{\beta_s \mu}(x,k)\,.
\label{eq:cov_dev_st}
\end{split}
\end{equation} 
Also, it is shown that given a metric, there is always a symmetric non-linear connection leading to the affine connections in spacetime making that the covariant derivative of the metric vanishes:
\begin{equation}
g_{\mu\nu;\rho}(x,k)\,=\,0\,.
\label{eq:covariant_derivative_2}
\end{equation}

In order to study the properties of the horizon, we need to know how the Lie derivative is deformed in this context. 
In~\cite{Barcaroli:2015xda,Relancio:2020zok} the modified Killing equation for a metric in the cotangent bundle was derived
\begin{equation}
\frac{\partial g_ {\mu\nu}(x,k)}{\partial x^\alpha} \chi^\alpha  -\frac{\partial g_ {\mu\nu}(x,k)}{\partial k_\alpha}\frac{\partial \chi^\gamma}{\partial x^\alpha}k_\gamma + g_{\alpha\nu}(x,k)\frac{\partial\chi^\alpha}{\partial x^\mu}+ g_{\alpha\mu}(x,k)\frac{\partial\chi^\alpha}{\partial x^\nu}\,=\,0\,,
\label{eq:killing}
\end{equation}
where $\chi^\alpha=\chi^\alpha(x)$ is momentum independent. In GR, where the metric does not depend on the momentum, the previous condition reduces as expected to the usual Killing equation
\begin{equation}
\chi_{\mu\,;\nu}+\chi_{\nu\,;\mu}\,=\,0\,.
\label{eq:killing_GR}
\end{equation}

\subsection{Deformed relativistic kinematics in curves spacetimes}
We summarize here our previous results about supplementing a deformed relativistic kinematics within a curved space-time. 

The deformed kinematics of DSR are usually obtained from Hopf algebras~\cite{Majid:1995qg}, being the $\kappa$-Poincaré kinematics~\cite{Majid1994}  the most studied example of this kind of construction. This kinematics has been understood from a geometrical point of view in~\cite{Carmona:2019fwf}. Given a de Sitter momentum metric $ \bar{g}$, translations can be used to define the associative deformed composition law, the Lorentz isometries lead to the Lorentz transformations, and the (squared of the) distance in momentum space is identified with the deformed Casimir.   

In~\cite{Relancio:2020zok,Pfeifer:2021tas}, we extended~\cite{Carmona:2019fwf} in order to consider in the same framework a deformed kinematics and a curved spacetime. For that aim, it is mandatory to consider the cotangent bundle geometry above discussed. The metric tensor $g_{\mu\nu}(x,k)$ in the cotangent bundle depending on space-time coordinates was constructed with the tetrad of spacetime and the original metric in momentum space, $\bar{g}$, explicitly
\begin{equation}
g_{\mu\nu}(x,k)\,=\,e^\alpha_\mu(x) \bar{g}_{\alpha\beta}(\bar{k})e^\beta_\nu(x)\,,
\label{eq:definition_metric_cotangent}
\end{equation}
where $\bar{k}_\alpha=\bar{e}^\nu_\alpha (x) k_\nu$, and $\bar{e}$ denotes the inverse of the tetrad of spacetime. 

In~\cite{Relancio:2020rys} it was also proved that this Hamiltonian can be identified with the square of the minimal geometric distance of a momentum $k$ from the origin of momentum space, measured by the momentum space length measure induced by the metric, relating  the Casimir and the metric in the following way~\cite{Relancio:2020zok}
\begin{equation}
\mathcal{C}(x,k)\,=\,\frac{1}{4} \frac{\partial \mathcal{C}(x,k)}{\partial k_\mu} g_{\mu\nu} (x,k) \frac{\partial \mathcal{C}(x,k)}{\partial k_\nu}\,.
\label{eq:casimir_metric}
\end{equation} 
A very important relation that the Casimir satisfies is that its delta derivative~\eqref{eq:delta_derivative} is zero, i.e.
\begin{equation}
\frac{\delta \mathcal{C}(x,k)}{\delta x^\mu}\,=\,0\,.
\label{eq:casimir_delta}
\end{equation} 
This is a necessary condition derived from the fact the Hamilton equations of motions are horizontal curves~\cite{Relancio:2020rys}. 

In~\cite{Pfeifer:2021tas} we showed that the most general form of the metric, in which the construction of a deformed kinematics in a curved space-time background is allowed, is a momentum basis whose Lorentz isometries are linear transformations in momenta, i.e., a metric of the form
\begin{equation}
\bar{g}_{\mu\nu}(k)\,=\,\eta_{\mu \nu} f_1 (k^2)+\frac{k_\mu k_\nu}{\Lambda^2} f_2(k^2)\,,
\label{eq:lorentz_metric}
\end{equation}
where $\Lambda$ is the high-energy scale parametrizing the momentum deformation of the metric and kinematics. From Eq.~\eqref{eq:definition_metric_cotangent} one obtains the following metric in the cotangent bundle when a curvature in spacetime is present
\begin{equation}
g_{\alpha\beta}(x,k)\,=\,a_{\alpha\beta}(x)f_1 (\bar{k}^2)+\frac{k_\alpha k_\beta}{\Lambda^2} f_2(\bar{k}^2)\,,
\label{eq:lorentz_metric_curved}
\end{equation}
where $a_{\mu \nu}(x) = e^\alpha_\mu(x) \eta_{\alpha\beta} e^\beta_\nu(x)$ is the GR metric. Therefore, one can use the definition of the space-time affine connection~\eqref{eq:affine_connection_st} to show that it is momentum independent, ending up being the same affine connection $ \Gamma^\rho_{\mu\nu}(x)$ of GR~\cite{Pfeifer:2021tas}.

Since we want the momentum metric~\eqref{eq:lorentz_metric} to be a de Sitter space (allowing us to define a deformed relativistic kinematics), a relationship between the functions $f_1$ and $f_2$ must hold. In particular, in~\cite{Relancio:2020rys} we found that, in order to be conserved the Einstein tensor defined in~\cite{miron2001geometry}, which is the same expression of GR, the metric~\eqref{eq:lorentz_metric_curved} must be conformally flat. Then, taking $f_2=0$ and imposing a momentum de Sitter space, one obtains from Eq.~\eqref{eq:lorentz_metric_curved} 
\begin{equation}
g_{\mu\nu}(x,k)\,=\,a_{\mu \nu}(x)\left(1-\frac{\bar{k}^2}{4\Lambda^2}\right)^2\,.
\label{eq:conformal_metric}
\end{equation}

\section{Killing equation in the cotangent bundle}
\label{sec:killing}
We shall now see that the above result concerning the cotangent bundle metric can be derived also on the base of simple requirements concerning the Killing equation~\eqref{eq:killing}. 

\subsection{Killing equation revisited}
In this subsection we consider a Schwarzschild black hole metric in some static (not time dependent) coordinates~\cite{Wald:1984rg}. The Killing vector is $\partial/\partial t$, as can be easily derived from Eq.~\eqref{eq:killing} when considering that the metric is momentum independent~\cite{Relancio:2020zok}, since the GR metric it is independent of time\footnote{Similar argument can be done for different space-time coordinates, since Eq.~\eqref{eq:killing} is invariant under diffeomorphisms, as can be seen for its construction~\cite{Relancio:2020zok}}. Then, as the metric in the cotangent bundle is constructed from the GR metric, this will be also independent of time, and therefore,  Eq.~\eqref{eq:killing} is automatically satisfied  for any metric of the form of~\eqref{eq:definition_metric_cotangent}, and in particular, for~\eqref{eq:lorentz_metric_curved}. 

In the GR case, Eq.~\eqref{eq:killing_GR} can be written as
\begin{equation}
{\chi^\rho}_{; \nu} a_{\rho\mu}(x)+{\chi^\rho}_{; \mu} a_{\rho\nu}(x) \,=\,0\,.
\label{eq:killing_GR2}
\end{equation}
Let us now require instead that Eq.~\eqref{eq:killing_GR} holds also for a generic cotangent bundle metric (of the form allowing for the connection with the consistent lift of a deformed kinematics to curved spacetimes of Eq.~\eqref{eq:lorentz_metric_curved}), i.e., that
\begin{equation}
{\chi^\rho}_{; \nu} g_{\rho\mu}(x,k)+{\chi^\rho}_{; \mu} g_{\rho\nu}(x,k) \,=\,0\,,
\label{eq:killing_DGR2}
\end{equation}
is satisfied. Following our previous discussion, both Eqs.~\eqref{eq:killing_GR2} and ~\eqref{eq:killing_DGR2} hold simultaneously. This will assure that all the surface gravity notions considered in the literature agree~\cite{Cropp:2013zxi}.

Let us now now make use of the explicit form of the metric~\eqref{eq:lorentz_metric_curved} to write Eq.~\eqref{eq:killing_GR2} as
\begin{equation}
\begin{split}
{\chi^\rho}_{; \nu}\left(f_1(\bar{k}^2) a_{\rho\mu}(x)+\frac{k_\rho k_\mu}{\Lambda^2}f_2(\bar{k}^2)\right)+{\chi^\rho}_{; \mu}\left(f_1(\bar{k}^2) a_{\rho\nu}(x)+\frac{k_\rho k_\nu}{\Lambda^2}f_2(\bar{k}^2)\right)\, =\,&\\
 {\chi^\rho}_{; \nu} \frac{k_\rho k_\mu}{\Lambda^2}f_2(\bar{k}^2) +{\chi^\rho}_{; \mu} \frac{k_\rho k_\nu}{\Lambda^2}f_2(\bar{k}^2)\,=\,&0\,,
\end{split}
\label{eq:killing_DGR3}
\end{equation}
where we have used Eq.~\eqref{eq:killing_GR2}. There are two possible ways in which the previous equation can be satisfied: either $f_2=0$ or 
\be
 {\chi^\rho}_{; \nu}  k_\rho k_\mu +{\chi^\rho}_{; \mu}  k_\rho k_\nu\,=\,0 \,.
 \label{eq:killing_DGR4}
\ee
Taking into account that the Killing vector $\chi^\mu$ is momentum independent, then its covariant derivative will also be (since as discussed above, the affine connection is the same one of GR). Therefore, we can derive Eq.~\eqref{eq:killing_DGR4} with  respect to the momentum two times, obtaining 
\be
 {\chi^\alpha}_{; \nu} \delta^\beta_\mu + {\chi^\beta}_{; \nu} \delta^\alpha_\mu+  {\chi^\alpha}_{; \mu} \delta^\beta_\nu + {\chi^\beta}_{; \mu} \delta^\alpha_\nu\,=\,0 \,.
\ee
For the considered Killing vector $\partial/\partial t$ this equation does not hold, which implies that the only way in which   Eq.~\eqref{eq:killing_DGR2} will be satisfied is by considering $f_2=0$. Imposing this condition on the metric~\eqref{eq:lorentz_metric_curved}  and   asking it to be a de Sitter space (so we are able to define a deformed relativistic kinematics as explained in the introduction) we fix the function $f_1$, which is precisely the one reproducing  Eq.~\eqref{eq:conformal_metric}.

\subsection{Killing equation in a conformally flat metric}

We shall now prove that in the momentum coordinates for which the metric takes the form of Eq.~\eqref{eq:conformal_metric}, Eq.~\eqref{eq:killing_DGR2} will be satisfied for every Killing vector.  
Let us assume that for the metric~\eqref{eq:conformal_metric}, Eq.~\eqref{eq:killing_GR} holds, which is tantamount  to saying 
\begin{equation}
\frac{\partial a_ {\mu\nu}(x)}{\partial x^\alpha} \chi^\alpha  + a_{\alpha\nu}(x)\frac{\partial\chi^\alpha}{\partial x^\mu}+ a_{\alpha\mu}(x)\frac{\partial\chi^\alpha}{\partial x^\nu} \,=\,0\,.
\label{eq:killing2}
\end{equation}

In order to prove our ansatz, we start by noticing that for the metric~\eqref{eq:conformal_metric} there is a simple relation between the $\delta$ derivative and the $\partial$ one.  As it was shown in~\cite{Relancio:2020rys}, the $\delta$ derivative of a function of $\bar{k}^2$ is zero if the affine connection is the one of GR, i.e.
\begin{equation}
\frac{\delta f(\bar{k}^2)}{\delta x^\mu}\,=\,0\,, \qquad \text{if} \qquad {H^\gamma}_{\rho\alpha}(x,k)\,=\, {\Gamma^\gamma}_{\rho\alpha}(x)
\label{eq:delta_k2}
\end{equation}
This is due to the fact that the $\delta$ derivative of the Casimir vanishes, as stated in Eq.~\eqref{eq:casimir_delta}, which also implies that the $\delta$ derivative of any function of the Casimir, that is, of any function of  $\bar{k}^2$, is zero. This implies that 
\begin{equation}
\frac{\delta g_{\mu\nu}(x,k)}{\delta x^\rho}\,=\,\left(1-\frac{\bar{k}^2}{4\Lambda^2}\right)^2 \frac{\partial a_{\mu\nu}(x)}{\partial x^\rho}\,.
\label{eq:delta_g}
\end{equation}

Therefore,  the first term of Eq.~\eqref{eq:killing} can be written as the sum of two terms
\begin{equation}
\frac{\partial g_ {\mu\nu}(x,k)}{\partial x^\alpha} \chi^\alpha \,=\,\left(\frac{\delta g_ {\mu\nu}(x,k)}{\delta x^\alpha} -\frac{\partial  g_ {\mu\nu}(x,k)}{\partial k_\rho}N_{\rho\alpha}(x,k)\right)\chi^\alpha\,=\,\left(\left(1-\frac{\bar{k}^2}{4\Lambda^2}\right)^2 \frac{\partial a_ {\mu\nu}(x)}{\partial x^\alpha}  -\frac{\partial  g_ {\mu\nu}(x,k)}{\partial k_\rho}N_{\rho\alpha}(x,k) \right)\chi^\alpha\,,
\end{equation}
where in the last step we have used Eq.~\eqref{eq:delta_g}. 

We can now expand the second term of the previous equation, obtaining
\begin{equation}
\begin{split}
 -\frac{\partial  g_ {\mu\nu}}{\partial k_\rho}N_{\rho\alpha}(x,k)\chi^\alpha\,&=\, -2 a_ {\mu\nu}(x)\left(1-\frac{\bar{k}^2}{4\Lambda^2}\right)\left(-\frac{1}{4\Lambda^2}\right) k_\sigma a^{\rho\sigma}(x) k_\gamma {\Gamma^\gamma}_{\rho\alpha}(x)\chi^\alpha \\
&=\, 2 a_ {\mu\nu}(x)\left(1-\frac{\bar{k}^2}{4\Lambda^2}\right)\left(\frac{1}{4\Lambda^2}\right) k_\sigma a^{\rho\sigma}(x) k_\gamma \frac{1}{2}a^{\gamma\delta}(x) \left(\frac{\partial a_{\delta\rho}(x)}{\partial x^\alpha}+\frac{\partial a_{\delta\alpha}(x)}{\partial x^\rho}-\frac{\partial a_{\alpha\rho}(x)}{\partial x^\delta}\right) \chi^\alpha  \\
&=\,  a_ {\mu\nu}(x)\left(1-\frac{\bar{k}^2}{4\Lambda^2}\right)\left(\frac{1}{4\Lambda^2}\right) k_\sigma a^{\rho\sigma}(x) k_\gamma \frac{\partial a_{\delta\rho}(x)}{\partial x^\alpha} \chi^\alpha\\
&=\, -2 a_ {\mu\nu}(x)\left(1-\frac{\bar{k}^2}{4\Lambda^2}\right)\left(\frac{1}{4\Lambda^2}\right) k_\sigma a^{\rho\sigma}(x) \frac{\partial \chi^\gamma}{\partial x^\alpha}k_\gamma\,=\,\frac{\partial g_ {\mu\nu}(x,k)}{\partial k_\alpha}\frac{\partial \chi^\gamma}{\partial x^\alpha}k_\gamma
\,,
\end{split}
\end{equation}
where in the first step we have used Eq.~\eqref{eq:nonlinear_connection}, in the second one the definition of the affine connection in GR,
\begin{equation}
{\Gamma^\rho}_{\mu\nu}(x)\,=\,\frac{1}{2}a^{\rho\sigma}(x)\left(\frac{\partial a_{\sigma\nu}(x)}{\partial x^\mu} +\frac{\partial a_{\sigma\mu}(x)}{\partial  x^\nu} -\frac{\partial a_{\mu\nu}(x)}{\partial x^\sigma} \right)\,,
\label{eq:affine_connection_GR}
\end{equation} 
in the third one the symmetry under the exchange of indexes $\delta \leftrightarrow \rho$, and in the forth one, Eq.~\eqref{eq:killing2} and the same symmetry. This expression can be written as a momentum derivative of the cotangent bundle metric, as we did in the first step. Therefore, since this term is the same one of the second of Eq.~\eqref{eq:killing} with opposite sign, we can write Eq.~\eqref{eq:killing} as
\begin{equation}
\left(1-\frac{\bar{k}^2}{4\Lambda^2}\right)^2 \left(\frac{\partial a_ {\mu\nu}(x)}{\partial x^\alpha} \chi^\alpha + a_{\alpha\nu}(x)\frac{\partial\chi^\alpha}{\partial x^\mu}+ a_{\alpha\mu}(x)\frac{\partial\chi^\alpha}{\partial x^\nu}\right)\,=\,0\,,
\label{eq:killing3}
\end{equation}
which is automatically true if Eq.~\eqref{eq:killing2} holds. Hence, with the particular choice of the momentum metric, Eq.~\eqref{eq:conformal_metric}, the standard Killing equation Eq.~\eqref{eq:killing_GR} is still satisfied.

\section{Main notions of surface gravity}
\label{sec:surface_gravity}
In GR, there are different definitions for the surface gravity~\cite{Cropp:2013zxi}. In this section, we compute the two main ones related respectively to the peeling off properties near the horizon and the inaffinity of null geodesics on the horizon.

\subsection{Peeling off properties of null geodesics}
Due to the form of the metric~\eqref{eq:lorentz_metric_curved}, the Casimir defined as the squared of the distance in momentum space as in Eq.~\eqref{eq:casimir_metric} is a function of $\bar{k}^2$. Then, for massless particles the same relationship of between energy and momentum holds in this deformed scenario. This also means that photons in this scenario follow the same trajectories of GR, implying an existence of an universal horizon at $2M$, independently of the energy of the particle.\footnote{
As we mentioned above, due to the form of the metric~\eqref{eq:lorentz_metric_curved}, the Casimir is a function of $\bar{k}^2$. This means that, on one hand there is not any modification of the dispersion relation for massless particles, and on the other, the modification of massive particles is of the order of $m^2/\Lambda^2$, being $m$ the mass of the particle, which is completely negligible. Therefore, in the ultraviolet regime in which particles scape from the horizon of the black hole, and then masses can be neglected, massive particles see the same horizon. This is a very important check of consistency since, as commented in the introduction, otherwise the black hole would be a perpetuum mobile~\cite{Dubovsky:2006vk}.} 

As it was shown in~\cite{Cropp:2013zxi}, the surface gravity can be defined as the peeling off of null geodesics
\begin{equation}
\frac{d |r_1(t)-r_2(t)| }{d t}\,\approx \, 2 \kappa_{\text{peeling}}(t) |r_1(t)-r_2(t)|\,,
\label{eq:surface_gravity}
\end{equation}
where $r_1(t)$ and $r_2(t)$ are two null geodesics on the same side of the horizon and the normalization of $\kappa_{\text{peeling}}$ is chosen so to coincide with $\kappa_\text{inaffinity}$ in the GR limit.

In order to compute it, we need to use some coordinates for which the final result does not diverge at the horizon. While there is no problem in GR, due to the momentum dependence of the metric~\eqref{eq:lorentz_metric_curved} some coordinates are not well behaved at the horizon. This means that, for example, Schwarzschild coordinates used in~\cite{Cropp:2013zxi} cannot be employed here. We will consider the Eddington-Finkelstein coordinates~\cite{Poisson:2009pwt} in the following
\begin{equation}
a_{vv}\,=\, -\left(1-\frac{2M}{r}\right)\,,\qquad a_{vr}\,=\,1\,,\qquad a_{rr}\,=\,0\,,\qquad a_{\theta\theta }\,=\,r^2\,,\qquad a_{\varphi\varphi}\,=\,r^2 \sin^2{\theta}\,.
\label{eq:ef_coordinates}
\end{equation}

This makes that radial component of the momentum is related to the zero component as
\begin{equation}
k_\mu a^{\mu\nu}(x)k_\nu\,=\,0\qquad\implies\qquad k_r\,=\, -\frac{2k_v}{1-2M/r}\,.
\end{equation}
Therefore, from Eq.~\eqref{eq:lorentz_metric_curved} one finds  
\begin{equation}
\frac{dr}{dt}\,=\,  \frac{1}{2}\left(1-\frac{2M}{r}\right)\,,
\label{eq:ef_velocity}
\end{equation}
which is independent of the energy. This result is an obvious outcome from the fact that the Casimir is underformed for massless particles. Hence, the same result of GR is obtained from Eq.~\eqref{eq:surface_gravity}
\begin{equation}
 \kappa_{\text{peeling}}\,=\,\frac{1}{4M}\,.
\end{equation}

This differs from the result obtained in~\cite{Relancio:2020zok} for the momentum metric corresponding to the bicrossproduct basis of $\kappa$-Poincaré~\cite{Gubitosi:2013rna,Carmona:2019fwf}. However, it is important to note that this basis cannot be consistently lifted to curved spacetimes from our proposal, as it was shown in~\cite{Pfeifer:2021tas}.

\subsection{Inaffinity of null geodesics}
Also, in~\cite{Cropp:2013zxi} it was shown that for a Killing vector $\chi^\mu$ one finds 
 \begin{equation}
 \chi^\nu\,  {\chi^\mu}_{;\nu} \,=\,\kappa_{\text{inaffinity}} \chi^\mu\,.
\label{eq:inaffinity}
\end{equation}
Again, we consider the Eddington-Frinklenstein coordinates. As commented before, it is easy to see from Eq.~\eqref{eq:killing} that, if the Killing vector of a metric in GR does not depend on the coordinates, it will be also constant in this framework, being in this case $\partial/\partial t$. Due to the fact that the space-time affine connection is the same one of GR, it is easy to obtain 
\begin{equation}
 \kappa_{\text{inaffinity}}\,=\,\frac{1}{4M}\,.
\end{equation}

\section{Killing equation and selection of momentum basis}
\label{sec:killing_basis}
In~\cite{Cropp:2013zxi}, several alternative definitions of surface gravity were discussed, showing that all of these coincide for Killing horizons. As explained previously, in our framework the same Killing vector of GR is present. However, the same framework implies a modification of the standard GR Killing equation Eq.~\eqref{eq:killing_GR}: the momentum dependent Eq.~\eqref{eq:killing}.
We shall show below that this reflects in an inequivalence of the other common definitions of surface gravity, which in general lead to a momentum dependent results different from what was found in the previous section. Nevertheless, Eq.~\eqref{eq:killing_GR} indeed holds for the momentum basis of Eq.~\eqref{eq:conformal_metric} (which we stress was derived by completely different arguments in~\cite{Relancio:2020rys}). Therefore, different definitions of surface gravity will again coincide in these particular momentum coordinates. This disambiguation of the definition of surface gravity lends then further support to the physical relevance of this particular momentum basis.

\section{Different notions of surface gravity}
\label{sec:different_notions}
We are going to check that other definitions of surface gravity coincides with the particular choice of the metric~\eqref{eq:conformal_metric}.

\subsection{Null normal derivative}
Another definition with respect to the previously discussed peeling and inaffinity ones is the null normal derivative evaluated in the horizon (see for example~\cite{Cropp:2013zxi})
\begin{equation}
\left(\chi^\nu \chi_\nu\right)_{;\mu}\,=\,-2 \kappa_{\text{normal}} \chi_\mu\,,
\end{equation}
which is equivalent to 
\begin{equation}
\chi^\nu \chi_{\nu;\mu}\,=\,- \kappa_{\text{normal}} \chi_\mu\,.
\label{eq:kappa_normal}
\end{equation}
Using Eq.~\eqref{eq:killing_GR} (as it holds for our metric~\eqref{eq:conformal_metric}), one can rewrite the previous expression as
\begin{equation}
\chi^\nu \chi_{\mu;\nu}\,=\, \kappa_{\text{normal}} \chi_\mu\,\qquad \implies \qquad \chi^\nu {\chi^\mu}_{;\nu}\,=\, \kappa_{\text{normal}} \chi^\mu\,,
\label{eq:kappa_normal2}
\end{equation}
from which we see that $\kappa_{\text{normal}}=\kappa_{\text{inaffinity}}$.

\subsection{Generator}
Another possible definition of the black hole surface gravity is the so-called $\kappa$-generator defined as~\cite{Cropp:2013zxi}  
\begin{equation}
 \kappa^2_{\text{generator}}\,=\,-\frac{1}{2}{\chi^\mu}_{;\sigma}{\chi^\nu}_{;\lambda}g_{\mu \nu}(x,k)g^{\sigma \lambda}(x,k)\,.
\end{equation}
As we have seen previously, the same Killing vector of GR is a Killing vector in this scheme. Then, the only momentum dependency arises from the metric. Due to the conformal form of the metric~\eqref{eq:conformal_metric}, from the previous equation one obtains 
\begin{equation}
 \kappa^2_{\text{generator}}\,=\,-\frac{1}{2}{\chi^\mu}_{;\sigma}{\chi^\nu}_{;\lambda}\left(1-\frac{\bar{k}^2}{4\Lambda^2}\right)^2 a_{\mu \nu}(x) \left(1-\frac{\bar{k}^2}{4\Lambda^2}\right)^{-2}a^{\sigma \lambda}(x)\,=\,-\frac{1}{2}{\chi^\mu}_{;\sigma}{\chi^\nu}_{;\lambda}a_{\mu \nu}(x)a^{\sigma \lambda}(x)\,.
\end{equation}
Hence, since this definition is equivalent to the others in GR~\cite{Wald:1984rg,Poisson:2009pwt}, it leads to the same value also in our scheme.

\subsection{Wick rotation}
A different way to obtain the surface gravity in GR is by the Wick rotation (explained in Ch.6 of~\cite{frolov2011introduction}). We firstly resume the case of GR using the Schwarzschild coordinates
\begin{equation}
a_{tt}(x)\,=\, -\left(1-\frac{2 M}{r}\right)\,,\qquad a_{rr}(x)\,=\, \left(1-\frac{2 M}{r}\right)^{-1}\,, \quad  a_{\theta\theta}(x)\,=\, r^2\,,\qquad  a_{\phi\phi}(x)\,=\, r^2 \sin^2(\theta) \,.
\label{eq:Schw_metric}
\end{equation}

We consider the t-r line element 
\begin{equation}
d\gamma^2\,=\,-dt^2  \left(1-\frac{2 M}{r}\right)+ dr^2 \left(1-\frac{2 M}{r}\right)^{-1}\,.
\end{equation}
We start by making a Wick rotation obtaining
\begin{equation}
d\gamma^2\,=\,dt_E^2  \left(1-\frac{2 M}{r}\right)+ dr^2 \left(1-\frac{2 M}{r}\right)^{-1} \,.
\end{equation}
We can define in the vicinity of the horizon the proper length distance from the horizon~\cite{frolov2011introduction}
\begin{equation}
\rho\,=\,\int^r_{2M}\frac{dr}{\sqrt{1-2M/r}}\,\implies\, 1-\frac{2 M}{r}\,=\,16M^2 \rho^2\,.
\end{equation}
Then, the Euclidean line element takes the following form
\begin{equation}
d\gamma^2\,=\,\left(16M^2\rho^2 dt_E^2 + d\rho^2 \right) \,.
\end{equation}
This line element will not be singular if the temporal coordinate behaves as an ``angular coordinate'' of the plane t-r. This implies that time must be periodical: 
\begin{equation}
d\gamma^2\,=\,\rho^2 \frac{dt_E}{\kappa^2}^2 + d\rho^2  \,,
\end{equation}
being $\kappa=1/4M$. 

It is obvious that the only way in which this procedure can be followed in our scheme is by using the conformally flat metric~\eqref{eq:conformal_metric}, obtaining then the same result of GR.

\section{Conclusion}
\label{sec:conclusions}

In this work we have studied different notions of surface gravity of a Schwarzschild black hole in a rainbow geometry in the DSR scenario. This study differs from previous works in the literature because here we have taken into account that, in order describe the kinematics of DSR, the momentum metric must be a maximally symmetric momentum space. In this way, the deformed kinematics are encoded in the geometrical ingredients of the cotangent bundle metric. 

As we have seen, both main notions of surface gravity lead to the same result of GR. However, the only way in which different definitions lead to the same momentum independent result is by selecting a conformally flat momentum metric, selecting a particular momentum basis of $\kappa$-Poincaré. This basis is the same one found in a previous work by imposing the conservation of the Einstein tensor.

This could seem to imply that the Hawking temperature is universal, independent of the energy of the emitted particle, when computed from the GR formula $T=\kappa/2\pi$. However, as discussed in the introduction, the formal derivation of the Hawking radiation requires a QFT in DSR, which at present is unknown.  

\section*{Acknowledgments}
JJR acknowledges support from the INFN Iniziativa Specifica GeoSymQFT. SL acknowledge funding from the Italian Ministry of Education and Scientific Research (MIUR) under the grant PRIN MIUR 2017-MB8AEZ. The authors would also like to thank support from the COST Action CA18108.

\end{document}